\documentclass[amsmath,amssymb,twocolumn,amsfonts,twocolumn]{revtex4}
\usepackage{graphicx}
\def \beq {\begin{equation}}
\def \eeq {\end{equation}}
\def \tr {\rm Tr}

\begin{document}
\title{Entropy Considerations in Spin-Selective Radical-Ion-Pair Reactions}
\author{Iannis K. Kominis}
%\email{ikominis@iesl.forth.gr}

\affiliation{Department of Physics, University of Crete, Heraklion
71103, Greece}

\begin{abstract}
Radical-ion-pair reactions were recently shown to manifest a host
of non-trivial quantum effects accounted for by quantum measurement
theory. An alternative approach purporting to describe the fundamental quantum dynamics of spin-selective radical-ion pair reactions was introduced most recently, bringing to three the competing theories, including the one traditionally used in spin chemistry. We here consider entropy as a fundamental concept enabling a comparison of the predictions of these theories against what is physically acceptable on quite general grounds.
\end{abstract}

\maketitle

Spin-selective radical-ion-pair reactions are at the core of spin
chemistry \cite{steiner}. It was recently shown \cite{kominis_PRE} that
radical-ion-pair reactions manifest non-trivial quantum effects
that were masked by the traditional treatment of these
reactions used since the advent of spin chemistry. Using the
specific Hamiltonian interactions pertaining in radical-ion pairs,
specifically taking into account electron tunnelling during charge
recombination, the density matrix equation describing
non-reacting radical-ion pairs was derived \cite{kominis_PRE}
based on basic quantum measurement theory considerations.
Most recently, yet another density matrix equation supposed to
describe radical-ion-pair reactions was suggested by Jones \& Hore
\cite{JH}, advocated to also follow from quantum measurement theory.
Moreover, two papers appeared \cite{purtov,ivanov} purporting to have derived from first principles the
traditional master equation of spin chemistry. 
We will here consider the purity of the spin state of radical-ion pairs as a guiding concept enabling a comparison of
all three theories against what is expected on general grounds, like the second law of thermodynamics. We will not address the traditional theory individually, since it suffers from the exact same problems as the Jones-Hore theory, problems to be outlined in the following. 

The fate of radical-ion-pair reactions is determined by the time
evolution of $\rho$, the density matrix describing the spin state
of the molecule's two electrons and any number of existing magnetic nuclei. In 2009 Kominis derived the Lindblad-type master 
equation
\beq
{{d\rho}\over {dt}}=-i[{\cal H},\rho]-{{k_{S}+k_{T}}\over 2}(Q_{S}\rho+\rho Q_{S}-2Q_{S}\rho Q_{S})\label{kom}
\eeq
where ${\cal H}$ is the magnetic interaction Hamiltonian inducing singlet-triplet mixing, $k_{S}$ and $k_{T}$ are the singlet and triplet recombination rates and $Q_{S}$ and $Q_{T}$ are the singlet and triplet projection operators (only $Q_{S}$ appears in \eqref{kom} because both the singlet and the triplet reservoir perform a continuous measurement of $Q_{S}$). The above trace-preserving master equation describes the density matrix evolution of {\it non-reacted} (i.e. not yet recombined) radical-ion pairs. It has been augmented by trace-decaying terms describing the recombination process \cite{kom_short}, but they are not needed for the present discussion. The master equation \eqref{kom} describes singlet-triplet decoherence induced by the continuous measurement of the singlet and triplet reservoirs. These reservoirs are the excited vibrational levels of the singlet and triplet neutral molecules resulting from the charge recombination, as will be illustrated in the following. 
 
A different approach was introduced by Jones \& Hore \cite{JH}, based on the
master equation  \begin{small}\begin{equation} {{d\rho}\over
{dt}}=-i[{\cal H},\rho]-(k_{S}+k_{T})\rho+k_{S}Q_{T}\rho
Q_{T}+k_{T}Q_{S}\rho Q_{S}\label{me_JH}\end{equation}\end{small}
We will here examine the repercussions of this theory regarding the purity of the spin state of radical-ion pairs. 
For simplicity, we will here consider an idealized radical-ion pair with no nuclear spins and at zero magnetic field. That is, we completely neglect singlet-triplet mixing induced by magnetic interactions. Furthermore, we consider just one recombination channel, e.g. the singlet, and thus set $k_{T}=0$. Finally, we take as initial state the coherent superposition $|\psi\rangle=(|S\rangle+|T\rangle)/\sqrt{2}$. So in this example the Jones-Hore master equation reads 
\beq
{{d\rho}\over
{dt}}=-k_{S}\rho+k_{S}Q_{T}\rho Q_{T}\label{me_JH}
\eeq
The philosophy of postulating such a master equation is the following \cite{JH}: if the density matrix of a single radical-ion pair at time $t$ is $\rho_{t}$, then there are three different scenarios for what can happen in the following time interval $dt$: (i) with probability $1-k_{S}dt$ nothing happens, keeping $\rho_{t+dt}=\rho_{t}$, (ii) with probability $k_{S}dt\tr\{Q_{S}\rho_{t}\}$ a singlet recombination takes place leading to $\rho_{t+dt}=0$, and (iii) with probability $k_{S}dt\tr\{Q_{T}\rho_{t}\}$ the molecule's state is projected to the triplet state,  making $\rho_{t+dt}=Q_{T}\rho_{t}Q_{T}/\tr\{Q_{T}\rho_{t}\}$. Averaging all three possibilities leads to \eqref{me_JH}.
\begin{figure}
\includegraphics[width=5.0 cm]{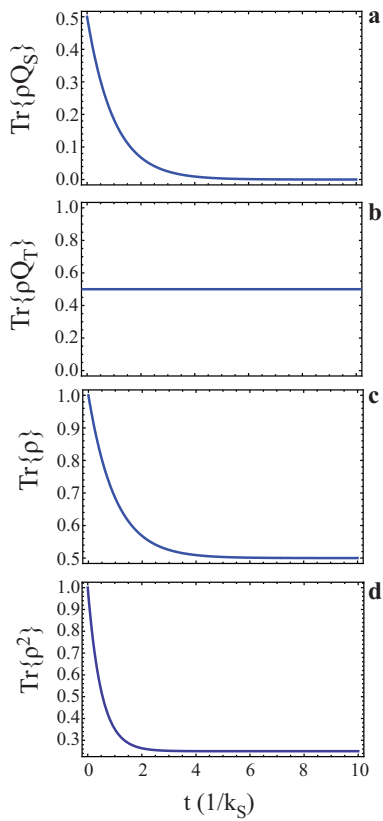}
\caption{Time evolution according to the Jones-Hore theory of (a) the singlet expectation value, (b) the triplet expectation value, (c) the number of existing radical-ion pairs, and (d) the purity of their quantum state, for the simple example of no singlet-triplet mixing, just the singlet recombination channel and a coherent initial state.} \label{f1}
\end{figure}
However, equation \eqref{me_JH} was not derived from the usual procedure of considering (i) the quantum system under study and (ii) the measurement-reservoir degrees of freedom, evolving the combined state under the specific system-reservoir coupling and then tracing out the reservoir degrees of freedom. It was postulated to follow from general considerations in analogy with the well known textbook example \cite{raimond} of a single two-level atom observed by a photodetector. If the atom is initially in a coherent state $(|e\rangle+|g\rangle)/\sqrt{2}$, and if the photodetector does not register a photon until time $t$, the state of the atom becomes $(e^{-\Gamma t/2}|e\rangle+|g\rangle)/{\cal N}$, where ${\cal N}$ is the normalization constant and $\Gamma$ the spontaneous decay rate. The interpretation is that not observing a photon, our knowledge of the atom's state acquires more ground-state "character". The Jones-Hore theory makes the seemingly innocent analogy that in the case of a radical-ion pair that reacts through the singlet channel, the state of a non-reacting radical-ion pair acquires more triplet-state "character". However, this analogy breaks down if examined more carefully. Before explaining why this is the case, we will present the solution of \eqref{me_JH} for the specific example we are considering.

\begin{figure}
\includegraphics[width=5.0 cm]{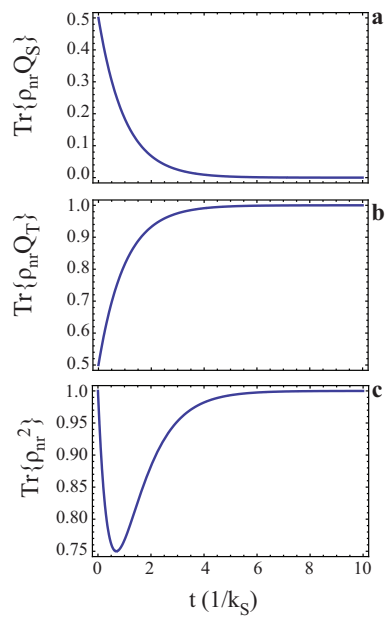}
\caption{Time evolution according to the Jones-Hore theory of (a) the singlet expectation value, (b) the triplet expectation value, and (c) the state purity of non-reacting radical-ion pairs.} \label{f2}
\end{figure}

In Fig.\ref{f1} we plot as a function of time (a) the expectation value $\langle Q_{S}\rangle=\tr\{Q_{S}\rho\}$, (b) the expectation value $\langle Q_{T}\rangle=\tr\{Q_{T}\rho\}$, (c) the normalization of the density matrix $\tr\{\rho\}$, which represents the number of surviving radical-ion pairs, and (d) the purity of the state $\tr\{\rho^{2}\}$. The interpretation of these results is that (a) half of the radical-ion pairs react through the singlet channel, (b) the other half remain locked in the non-reacting triplet state, (c) the number of radical ion pairs is, as expected,  reduced by half as $t\rightarrow\infty$, and (d) we start with a pure state and end up with a pure (triplet) state. The latter conclusion can be better visualized by considering non-reacting radical-ion pairs. To that end, we normalize $\langle Q_{S}\rangle$ and $\langle Q_{T}\rangle$ with $\tr\{\rho\}$ whereas we normalize the state purity $\tr\{\rho^{2}\}$ with  $\tr\{\rho\}^{2}$. The results are shown in Fig.\ref{f2}. The interpretation is that as long as a radical-ion pair does not react, its state approaches the pure triplet. From the construction of the Jones-Hore master equation \eqref{me_JH}, in the example at hand there are the following two possibilities for a non-reacting evolution between $t$ and $t+dt$: (i) $\rho_{t+dt}=\rho_{t}$, or (ii) $\rho_{t+dt}=|T\rangle\langle T|$. The average of all possible trajectories embodying these two possibilities is what is plotted in Fig.\ref{f2}. So we start from a pure state, this becomes mixed, and then it returns to be pure again. At first glance this result is intuitive and acceptable. It {\it seems to be} analogous to a situation where a decaying atom is initially in the pure excited state, it becomes entangled with the radiation field, hence the atomic state is mixed, and then it decays to the ground state, becoming pure again. Or, if we have an ensemble of spins initially in the coherent state $(|\uparrow\rangle+|\downarrow\rangle)/\sqrt{2}$, and they undergo a $T_2$ process, the state becomes mixed, $|\uparrow\rangle\langle\uparrow|/2+|\downarrow\rangle\langle\downarrow|/2$. If a $T_1$ process follows, we would end up with a pure spin state again.
However, here there is a problem. Among the $N$ molecules, $N/2$ non-reacting ones will evolve in trajectories that average to Fig.\ref{f2}. So for the ensemble of {\it non-reacting radical-ion pairs} we have the entropy initially increasing ($\tr\{\rho^{2}\}$ falls), and then decreasing ($\tr\{\rho^{2}\}$ rises back to 1). What makes up for this entropy decrease? The molecules are freely evolving, i.e. there is no energy exchange nor any entropy exchange with the environment. These molecules {\it do not react}, so there is nothing to take up the entropy loss, {\it as in the case of decaying atoms or spins}. For example, when atoms spontaneously decay to the ground state, the entropy decrease mentioned above is counterbalanced by the entropy increase of the radiation field. Similarly, when the spins undergo a $T_{1}$ process and end up in a pure state after having been mixed by a $T_2$ process, the entropy is taken up by the reservoir responsible for the $T_1$ process. How is the loss of entropy of non-reacting radical-ion pairs balanced?

\begin{figure}
\includegraphics[width=6.0 cm]{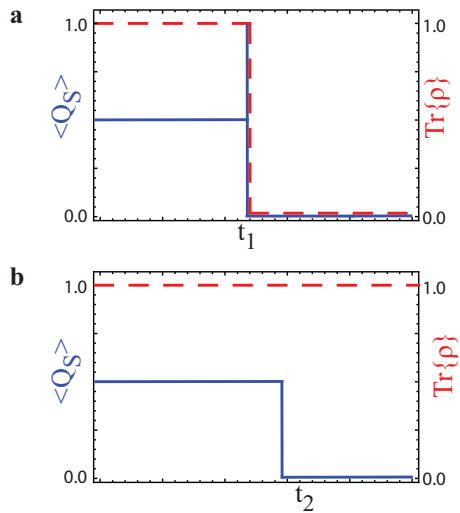}
\caption{Example of a quantum trajectories according to the Jones-Hore theory. (a) A reacting radical-ion pair according to the Jones-Hore theory. The expectation value of $Q_{S}$ jumps to one and at the same time $t_1$ the radical-ion pair singlet-recombines. (b) A non-reacting trajectory. At some random time $t_2$ the radical pair's state jumps to the triplet state and stays there forever. Blue solid line is the expectation value $\langle Q_{S}\rangle$, red dashed line is $\tr\{\rho\}$.} \label{trajJH}
\end{figure}

\begin{figure}
\includegraphics[width=6.0 cm]{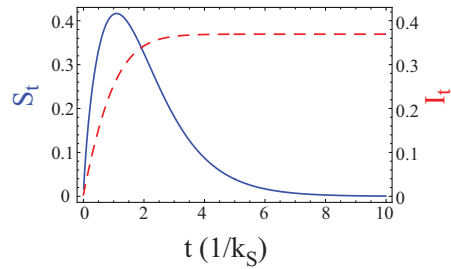}
\caption{Von Neumann entropy of non-reacting radical-ion pairs according to the Jones-Hore theory (blue solid line) and information gained by the reacting radical-ion pairs (red dashed line). The latter cannot accommodate the entropy loss taking place after $t\approx 1/k_{S}$. } \label{inf}
\end{figure}

To illustrate these considerations in more detail, we consider single quantum trajectories of reacting and non-reacting radical-ion pairs according to the Jones-Hore theory. As mentioned previously, the density matrix $\rho_{t}$ of a radical-ion pair at time $t$ can evolve in either of the three ways (i) $\rho_{t+dt}=\rho_{t}$, (ii) $\rho_{t+dt}=|T\rangle\langle T|$ and (iii) $\rho_{t+dt}=0$. 
Let us first look at the evolution of a radical-ion pair that does react at some point, i.e. at some random time possibility (iii) is realized. Of course, until then $\rho_{t}=\rho_{0}$, i.e. possibility (i) is realized. An example of such a trajectory is shown in Fig.\ref{trajJH}a. At time $t_1$ the radical pair's state jumps to the singlet state and recombines {\it at the same instant}.
An example of a non-reacting trajectory is shown in Fig.\ref{trajJH}b. At some random time $t_2$, $\langle Q_{S}\rangle$ jumps to $\langle Q_{S}\rangle=0$ ($\langle Q_{T}\rangle=1$) and the state is locked there forever. So the average of non-reacting trajectories up to time $t$ includes trajectories that have remained at the initial state $\rho_{0}$ up to time $t$ and trajectories that have been projected to the triplet state at some earlier time. That's where the decreasing purity of the non-reacting trajectories stems from. According to the Jones-Hore theory, the weight of the former decays as $e^{-k_{S}t}$, whereas the weight of the latter approaches 1/2 as $(1-e^{-k_{S}t})/2$. That's why as $t\rightarrow\infty$ the purity of the state of non-reacting trajectories approaches 1. But what after all compensates for the unphysical spontaneous entropy loss of the non-reacting radical-ion pairs? Is it the information gained from the reacting molecules? Not really. In Fig.\ref{inf} we plot (i) the von-Neumann entropy of $\rho_{nr}$, the density matrix of the non-reacting-radical ion pairs, which is $S_{t}=-\tr\{\rho_{nr}\ln\rho_{nr}\}$, and (ii) the information gained up to time $t$ from the reacting molecules. This is easily calculated as follows: Let $p_{S}=\tr\{Q_{S}\rho_{nr}\}$ be the singlet probability at time $t$ of a radical-ion pair just before it reacts. 
The information content of this molecule's state is $s_{I}=-p_{S}\ln p_{S}-(1-p_{S})\ln(1-p_{S})$. When the radical-ion pair reacts we have $p_{S}=1$, hence the information gain is $s_{I}$ itself. There are $k_{S}dt\tr\{Q_{S}\rho\}$ molecules reacting in the time interval between $t$ and $t+dt$, so the total information gain up to time $t$ is $I_{t}=\int_{0}^{t}{dtk_{S}\tr\{Q_{S}\rho\}s_{I}}$. From Fig.\ref{inf} it is seen that from $t\approx 1/k_{S}$ and on, the von-Neumann entropy of $\rho_{nr}$ drops from 0.4 to zero, but the information gain is just about 0.1, i.e. it cannot accommodate for this entropy loss.  

\begin{figure}
\includegraphics[width=5.0 cm]{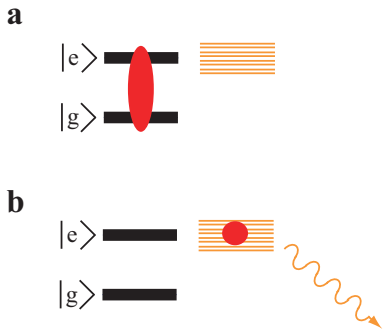}
\caption{(a) Two-level atom initially in a coherent state $(|e\rangle+|g\rangle)/\sqrt{2}$. The radiation field reservoir is coupled to the excited state $|e\rangle$. (b) If the atom decays, a photon occupies one of the radiation reservoir states. This is definitely observable. If the atom does not decay, then the probability that it is in the ground state increases.} \label{atom}
\end{figure}

\begin{figure}
\includegraphics[width=8.5 cm]{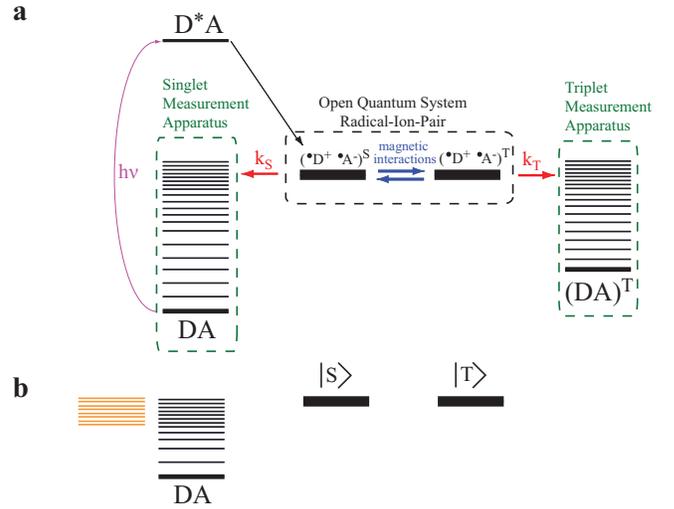}
\caption{(a) Radical-ion pair quantum dynamics. A photon excites the singlet neutral precursor molecule DA into D$^*$A, and a charge transfer creates the radical-ion pair.
The excited vibrational levels (DA)$^*$ of the neutral DA molecule form the measurement reservoir, which (i) acts as a measurement device for the radical pair's spin state, and (ii) acts as a sink of radical-ion pairs, i.e. in the event of recombination, the electron tunnels into a reservoir state and a fast spontaneous decay results in the ground state DA (which is the singlet product) and a photon emission. Similar for the triplet reservoir. (b) A simplified version of (a) in the case of just a singlet recombination channel. The (orange) reservoir on the left of the singlet reservoir depicts the radiation reservoir (or some other phonon reservoir) to which the excited vibrational states of the singlet reservoir are coupled.} \label{rip}
\end{figure}

We will now analyze the root of this shortcoming of the Jones-Hore theory. We will show where the analogy with the single atom mentioned above fails. 
In Fig.\ref{atom} we pictorially depict the atom's two energy levels and the radiation reservoir coupled to the excited state. An atom initially in a coherent superposition of the ground and excited states has the possibility of decaying. A photon will thus occupy one of the states of the radiation reservoir coupled to $|e\rangle$. If no photon is observed up to time $t$, one can claim that the atom's state has more $|g\rangle$ character. There is no other possibility. In other words, if a single occupation leaves the atomic system (spanned by $|g\rangle$ and $|e\rangle$) and goes to the radiation reservoir, then this event {\it is definitely observable in the outside world as a photon}. From the non-observation of a photon an {\it unambiguous} inference about the atom's state can be made. 
 However, the situation is not the same with radical-ion pairs. In Fig.\ref{rip}a we show the energy levels of a radical-ion pair and in Fig.\ref{rip}b we simplify this picture for use in the following. Evidently, we here have two reservoirs. One is the reservoir of the excited vibrational states of the neutral molecule. There is a second one. Each of those excited states is coupled to a radiation reservoir (or another phonon reservoir, it does not matter). 
 
\begin{figure}
\includegraphics[width=8.0 cm]{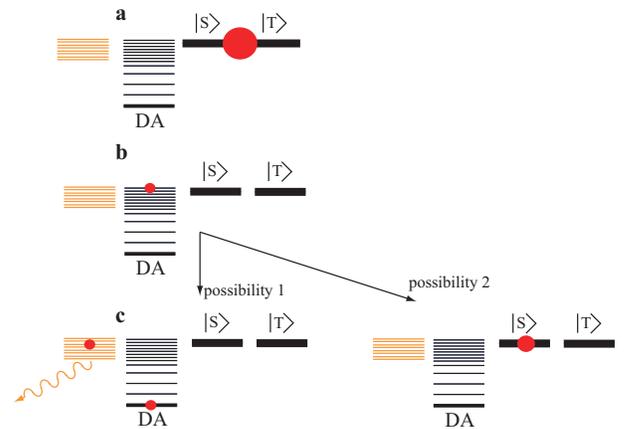}
\caption{Radical-ion pair (a) initially in the coherent singlet-triplet state. (b) The electron might tunnel to a singlet reservoir state, from which will follow (c) either a photon emission and the  production of a singlet neutral product i.e. the molecule DA (possibility 1), or it might tunnel back (possibility 2) to the singlet radical-ion pair state (virtual process) commencing again the spin-state evolution.} \label{rip1}
\end{figure}

\begin{figure}
\includegraphics[width=5.0 cm]{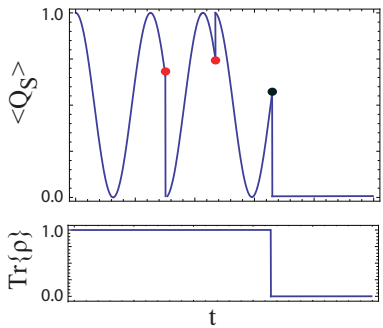}
\caption{A possible quantum trajectory of a radical-ion pair starting out from the singlet state, including singlet-triplet mixing for illustrative purposes. The two red circles are the times where we have a measurement outcome of $Q_{S}=0$ in the first and $Q_{S}=1$ in the second. Finally, at some random time (black circle) the radical-ion pair recombines, and from that point on $Q_{S}=Q_{T}=0$.} \label{traj}
\end{figure}

\begin{figure}
\includegraphics[width=6.0 cm]{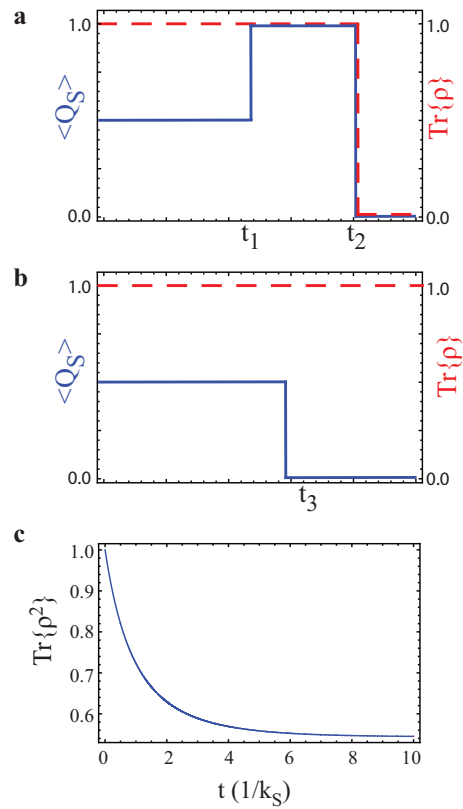}
\caption{Quantum trajectory of (a) a reacting radical-ion in which a measurement outcome of $Q_{S}=1$ takes place at time $t_1$ and the recombination reaction at a later time $t_2$. This trajectory is non-reacting until time $t_2$. There is nothing against $t_{2}>t_{1}$, and this fact is what the Jones-Hore theory fails to account for. (b) a non-reacting in which a measurement outcome of $Q_{S}=0$ takes place at time $t_3$. Blue solid line is the expectation value $\langle Q_{S}\rangle$, red dashed line is $\tr\{\rho\}$. (c) the average of all non-reacting trajectories produces the decaying purity resulting from \eqref{kom} for the example under study.} \label{trajrnr}
\end{figure}

How is the event of the charge recombination transmitted to the outside world? It is by the emission of the photon by the excited vibrational level and the creation of the singlet neutral product DA. If the radical-ion pair is at time $t$ in the coherent superposition $(|S\rangle+|T\rangle)/\sqrt{2}$ or in any state for that matter, and no photon is detected, i.e. no reaction takes place within the following time interval $dt$, the {\it only inference to be made} is that the singlet reservoir state was not occupied in this time interval. {\it No inference can be made about the radical-ion pair spin state}.
It is exactly this point where the analogy with the atom's case fails. 
In other words, the basis of the Jones-Hore theory is the erroneous association of the quantum measurement going on internally in radical-ion pairs due to the internal spin-state-measurement dynamics with the recombination event itself. To clarify this statement, in Fig.\ref{rip1}a we depict a radical-ion pair initially in the coherent state $(|S\rangle+|T\rangle)/\sqrt{2}$, i.e. the occupation resides in the quantum system spanned by $|S\rangle$ and $|T\rangle$. If the occupation leaves the system and goes to the singlet reservoir, this is {\it not observable} (Fig.\ref{rip1}b). What is observable is the next possible step (Fig.\ref{rip1}c) , i.e. the decay of the reservoir state and the creation of a photon and the ground state DA. But this is just one possibility. There is a second possibility (Fig.\ref{rip1}c). The occupation tunnels back to $|S\rangle$ and the spin state evolution continues on from there. 
This possibility is described by the second-order perturbation theory from which follows the master equation \eqref{kom} derived in \cite{kominis_PRE}. It is these virtual processes that lead to singlet-triplet decoherence introduced in \cite{kominis_PRE}. In other words, during a possible trajectory of a single radical-ion pair, one can have many instances where the continuous measurement induced by the singlet reservoir (and the triplet reservoir when it exists) results in $Q_{S}=1$ or $Q_{S}=0$. The Jones-Hore theory wrongly associates the former possibility with a concomitant singlet recombination. If the triplet recombination channel was also open, then the Jones-Hore theory would wrongly associate both the singlet ($Q_{S}=1$) and the triplet ($Q_{S}=0$) measurement outcomes with the singlet and triplet recombination event. The actual charge recombination event is, however, completely independent and can take place at any time. This is shown schematically for a more general example than the one considered here in Fig.\ref{traj}. 
The kinks produced in a smooth state evolution by the jumps to $Q_{S}=1$ or $Q_{T}=1$ are responsible for singlet-triplet decoherence. That is, the average of all trajectories like the one in Fig.\ref{traj} without a recombination event reproduces the master equation \eqref{kom} derived in \cite{kominis_PRE}. If we calculate the state purity according to this master equation, we find that it is a decreasing function of time. In the particular example we are considering here, a possible quantum trajectory for a reacting radical-ion pair is shown in Fig.\ref{trajrnr}a.
The radical-pair's spin state jumps to the singlet at time $t_1$ (measurement outcome $Q_{S}=1$), from which it recombines at time $t_2$. There is nothing against $t_{2}>t_{1}$, and this fact is what the Jones-Hore theory fails to account for.
An example of a non-reacting trajectory is shown in Fig.\ref{trajrnr}b. At some random instant $t_3$, the measurement outcome is $Q_{S}=0$, and the state jumps to the pure triplet, where it remains forever. We stress that up to time $t_2$ both trajectories are non-reacting trajectories. The average of all such pure non-reacting trajectories reproduces the singlet-triplet decoherence embodied in \eqref{kom}, which results in a monotonically decreasing state purity, as shown in Fig.\ref{trajrnr}c. In other words, in this particular example the infinite-time state of non-reacting trajectories is $\rho_{\infty}={1\over 2}|S\rangle\langle S|+{1\over 2}|T\rangle\langle T|$ as opposed to $\rho_{\infty}=|T\rangle\langle T|$ of the Jones-Hore theory. In our approach, it is easy to see that the von-Neumann entropy of non-reacting trajectories monotonically increases. Hence the problem facing the Jones-Hore theory does not exist. 

We can recapitulate the following consistent picture of what really happens. Up to some time $t$, non-reacting trajectories involve pure-state trajectories of molecules that have been projected to the triplet state, and pure-state trajectories of molecules that have been projected to the singlet state at some time earlier than $t$ and have not yet recombined. Until the latter do recombine, we have a mixed state of increased entropy. After they do, we end up with a pure triplet state. This entropy loss is counter-balanced by the entropy increase in the final photon/phonon reservoir. {\it It is through the reacting molecules that entropy is transferred from the radical-ion-pair system to the environment}.  In contrast, the Jones-Hore theory ignores the presence of the singlet vibrational-reservoir states of the radical-ion pair. This reservoir is in between the quantum system (radical-ion pair) and the outside world (photon reservoir). What goes on in this singlet reservoir is unobservable. Hence any information or the lack thereof in the outside world cannot be associated with the quantum system's state. However, the Jones-Hore theory does exactly that. Hence in this description, the non-reacting trajectories involve pure state-trajectories still in their initial state, and pure state trajectories that have been projected to the triplet state based on the non-observation of a reaction. The latter is, as explained before, an ambiguous inference on the radical-pair's spins state, and that's why it results in an unphysical entropy loss of {\it non-reacting trajectories}.
The difference between our consistent physical picture and the Jones-Hore theory is not just a philosophical curiosity. It has measurable consequences on the reaction time. If we define the latter to be the time at which the radical-ion pair population has decayed to a specified fraction of its original value, then the two approaches will predict different reaction times, exactly because the Jones-Hore theory makes the association "measurement outcome=recombination". We will further analyze reaction times in another manuscript.

I would like to acknowledge several fruitful discussions with Prof. Ulrich Steiner.

\end{document}